\documentclass[fleqn]{an}
\usepackage{graphicx}
\usepackage{times}
\usepackage{amssymb}
\usepackage{amsmath}
\usepackage{amsfonts}
\Volume{00}              
\Year{0000}              
\Month{00}               
\Pagespan{000}{000}      

\begin{document}
\lhead[\thepage]{L.V. Verozub: Sgr A* as probe  of the theory of
supermassive compact objects without event horizon }
\rhead[Astron. Nachr./AN~{\bf  XXX} (200X) X]{\thepage}
\headnote{Astron. Nachr./AN {\bf 32X} (200X) X, XXX--XXX}

\title{Sgr A* as probe  of the theory of supermassive compact
objects without  event horizon }

\author{L. Verozub}
\institute{Kharkov National University, Kharkov, Ukraine}
\date{Received {date will be inserted by the editor};
accepted {date will be inserted by the editor}}

\abstract{In the present paper  some consequences of  the
hypothesis  that the supermassive compact object in the Galaxy
centre relates to a class of objects without  event horizon are examined.
The possibility of the existence of such objects  was substantiated by the
author earlier. It is shown that  accretion of a surrounding
gas can cause   nuclear combustion in the surface
layer which, as a result of comptonization of the superincumbent
hotter layer, may  give a contribution to the observed Sgr A* radiation
 in the range $10^{15} \div 10^{20}\, Hz$.
  It is found a contribution of the possible proper
magnetic moment of the object to the observed synchrotron
radiation on the basis of   Boltzmann's equation  for
photons  which takes into account the influence of gravity to their
motion and frequency. We arrive at the conclusion that the
hypothesis of the existence in the Galaxy centre  of the
object with such extraordinary gravitational properties  at least
does not contradict observations. 
\keywords{dense matter, black hole physics, gravitation, accretion, radiation mechanism: non-thermal}
}

\correspondence{verozub@t-online.de;\\  leonid.v.verozub@univer.kharkov.ua}

\maketitle

\section{Introduction}
An analysis of stars motion in the Galaxy centre 
gives strong evidence for the existence here of a compact object with a
 mass of ($3\div4)\,10^{6}M_{\odot}$  associated with the radio source
Sgr A* (\cite{gensel1}; \cite{ gensel2}; \cite{ghez2};  \cite{ ghez1}). 
There are three
kind of an explanation of  observed
peculiarities of the object emission:\\
 --    gas accretion onto the central object - a supermassive
black hole (SMBH) ( \cite{melia};  \cite{ narayan}) ,\\
 --     ejection of  magnetised plasma from an environment of the
Schwarzschild  radius of the SMBH (\cite{falcke}; \cite{melia1}),\\
 -- explanations, based on  hypotheses of other nature of
the central object ( clusters of dark objects (\cite{maoz})  ,  a
fermion ball  (\cite{viollier}), boson stars (\cite{torres}; \cite{yuan1}).\\   

In the present paper   some consequences of the
assumption that   emission of Sgr A*  caused by the existence in
 the Galaxy  centre   of a supermassive compact object without 
event horizon  are examined.
 The existence of such  stable configurations of the
degenerated Fermi-gas of  masses $10^{2}\div10^{10}$ $M_{\odot}$,
 radiuses of which are  smaller than the Schwarzschild radius
$r_\mathrm{g}$,  is one of  consequences (\cite{verozub95})
 of   the metric-field equations
of gravitation  (\cite{verozub91}; \cite{verozub01}). According to  
the  physical principles founding the equations,    gravity of a compact object  manifests itself 
as a field  in     Minkowski space-time for a remote observer  in an inertial frame of reference,    
but become apparent 
as a space-time curvature   for the observer in  comoving  reference   frames associated with particles freely moving   in this field .
If the distance from a point  attractive mass is many larger than $  r_\mathrm{g}$ , 
the
physical consequences, resulting from these equations, are very close to  General Relativity results.
However, they are principally  other at  short distances from the central mass .
 The spherically- symmetric solution of these
equations in  Minkowski space-time   have no  event
horizon  and  physical singularity in the centre.

Since these gravitational equations were successfully tested by
 post-Newtonian effects in  the solar system and by  the binary pulsar
PSR 1913+16 (\cite{verkoch00}), and the stability of the
supermassive configurations predicted by them was 
sufficiently strictly  substantiated  (\cite{verkoch01}), it is of interest to
investigate the possibility of the existence of such  a  kind of an
object in the  Galaxy centre as  an alternative to the hypothesis
of  a supermassive black hole.

 The gravitational force of a point mass $M$ affecting a
free- falling particle  of mass $m$ is given by (\cite{verozub91}).
\begin{equation}
 F=-m\left[c^{2} C^{\prime}/2A +(A^{\prime}/2A  - 2 C^{\prime}/2C)  \overset{\cdot}{r}^{2}\right]  ,
\label{gravaccel1}%
\end{equation}
where
\begin{equation}
A=f^{\prime2}/C,\ C=1-r_\mathrm{g}/f,\ f=(r_\mathrm{g}^{3}+r^{3})^{1/3}.
 \label{ABC}
\end{equation}
In this equation $r$ is the radial distance from the centre,
$r_\mathrm{g} =2GM/c^{2}$,  $G$ is the
gravitational constant, $c$ is the  speed of light at infinity, the prime denotes the
 derivative with respect to $r$. 
 
 For   particles at rest ($\overset{\cdot}{r} =0$ ) 
\begin{equation}
F=-\frac{GmM}{r^{2}} \left[
1-\frac{r_\mathrm{g}}{(r^{3}+r_\mathrm{g}^{3})^{1/3}} \right]
\label{ForceStat}%
\end{equation}

\begin{figure}[h]
\resizebox{\hsize}{!} {\includegraphics[]{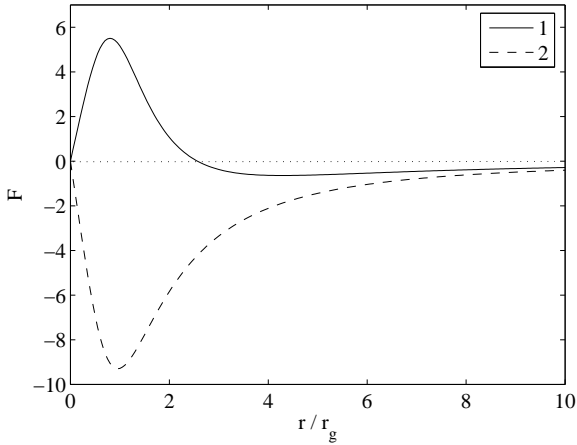}}
 \caption{The gravitational force (arbitrary units) affecting  free-falling  particles  (curve
1) and   particles  at rest (curve 2) near a point  attractive
mass $M$.}
 \label{fig: GRForce}
\end{figure}

 Fig. \ref{fig: GRForce} shows the force $F$ 
affecting    particles at rest  and  particles, free
falling from infinity with zero initial speed,   as the
function of the distance $\overline{r}=r/r_\mathrm{g}$ from the centre.

It follows from  Fig. \ref{fig: GRForce}  that the gravitational force affecting
free falling particles  
 changes its sign at  $r \approx 2  r_\mathrm{g}$ . Although
we still never observed particles motion  at  distances of the order of 
$r_\mathrm{g}$,   we can verify this result for very remote objects
in the Universe -- at  large cosmological redshifts .
Indeed, consider a simple model -  homogeneous  selfgravitating  dust-balls
  of  different-size  radiuses $r_\mathrm{b}$. 
 The force  acting on  particles at the  surface of such a ball     is given by Eq.
(\ref{gravaccel1})  where, in this case,  $r_\mathrm{g}=(8/3) \pi c^{-2} G \varrho_{b}  r_\mathrm{b}^{3}$ is the  Schwarzschild radius of  the
ball and $\varrho_{b}$ is  its density.
If  the ball density  is of the order of the observed density of  matter in the Universe ($\sim 10^{-29}  \mathrm{g} \, \mathrm{cm}^{-3}$), and
the its radius   $r_\mathrm{b}$   is not less than the radius of the  observed matter  ( $\sim  10^{28} \mathrm{cm}$    ), then
 $r_\mathrm{b} \leq r_\mathrm{g}$. It follows from       Fig. \ref{fig: GRForce}  that under the sircumstances 
    the radial accelerations   $F/m$   of  speckle particles  on the ball surface
 are positive.  The repulsive force give rise a deceleration of  the  expansion of such a selfgravitating   dust-like mass
 (  \cite{verozubA02} ).
  More accurate calculations at the redshift  $0<z \leq 1$       (\cite{verkoch3}) give a good accordance with  observation data  (\cite{riess})     .
 
We assume here for calculations that the mass $M$ of the central object is about  $3\,10^{6} M_{ \odot}$.
The  radius  $R$  of the object can be found   from the distribution of the density $\rho_\mathrm{m}$ of  matter inside the object. 
It can  be obtained by   solving  the  equations of the  hydrodynamical equilibrium: 
\begin{eqnarray}
\label{EquilibrHydEq}
dp_\mathrm{m}/dr=\rho_\mathrm{m} g_\mathrm{m},  \,\, d M_\mathrm{r} /dr=4 \pi r^2 \rho _\mathrm{m},\, \,   \\
 p_\mathrm{m}=p_\mathrm{m}(\rho_\mathrm{m}).  \nonumber
\end{eqnarray} 
 In these equations  $M_\mathrm{r}$ is the  mass  of matter  inside  the sphere $\mathcal{O}_\mathrm{r}$  of the radius $r$, $g_\mathrm{m}=F/m$  is the force   (\ref{ForceStat}) per unit of mass,  $r_\mathrm{g}=2 G \, M_\mathrm{r}  /c^2$ is the Schwarzschild radius of  the sphere $\mathcal{O}_\mathrm{r}$  and $p=p(\rho_\mathrm{m})$ 
 is the equation of state of   matter  inside  the object. The equation of state,  which is 
 valid for  the  density range  of 
   $(8 \div 10^{16} ) \,   \mathrm{g}   \,   \mathrm{cm}^{-3}$, is given by
 (\cite{harrison})
 \begin{equation}
p_\mathrm{m}= \left(   \frac{n_\mathrm{m}}{\partial n_\mathrm{m} /  \partial \rho_\mathrm{m} }    - \rho_\mathrm{m}  \right)  \, c^{2},
\end{equation} 
 where in CGS units
 \begin{equation}
n_\mathrm{m} = Q_{1} \rho_\mathrm{m} \left(    1+Q_{2} \rho_\mathrm{m}^{9/16}\right) ^{-4/9},
\end{equation} 
 $Q_{1}=6.0228 \,10^{23}$  and  $Q_{2}=7.7483 \,10^{-10}$
 . For the  gravitational force under consideration there are 
 two types of the solutions of  Eqs. ( \ref{EquilibrHydEq} ).  Besides the solutions describing white dwarfs and newtron stars,
  there are     solutions  with very large masses. For the mass of $ 3\,10^{6} M_{\odot}$ 
  the object radius resulting from Eqs.  (\ref{EquilibrHydEq} )   is about
 $0.4 R_{ \odot}=0.034\,  r_\mathrm{g}=3 \,10^{10} \mathrm{cm}   $ where $r_\mathrm{g}$ is the Schwarzschild radius  of the object. 

 It seems  at first glance   that  accretion onto the object with a solid surface must
lead to too  large  energy release  which
contradicts a comparatively low bolometric luminosity (
$\sim 10^{36}$ $\mathrm{erg}\ \mathrm{s}^{-1}$ ) of Sgr A*. However, it must be
taken into account that 
 the radial velocity of test particles free falling from infinity   is given by the equation (\cite{verozub91})
 \begin{equation}
v_\mathrm{ff}=c\left[ \dfrac{C}{A} (1-C)    \right]^{1/2}.
\label{vff}
\end{equation} 
 The velocity   fast decreases at $r < r_\mathrm{g}$. As a result, the velocity  at the surface is only about 
$3.3 \,10^{8}\,\mathrm{cm}\, \mathrm{s}^{-1}$ . Consequently, 
 at the accretion rate $\overset{\cdot}{M}=10^{-7}$
$M_{\odot}\ \mathrm{yr}^{-1}$ the amount of the released energy
$\overset{\cdot}{M}v^{2}/2$  is  equal to  
  $ 3.4 \,10^{35}\,   \mathrm{erg}\, \mathrm{s}^{-1}$.

\section{Atmosphere}
There are reasons to  believe  that the rate of  gas accretion onto
the supermassive object in the Galaxy centre due to  star winds
from  surrounding young stars is of the order of $10^{-7}$
$M_{\odot}\ \mathrm{yr}^{-1}$ (\cite{coker}).  Therefore, if the object has a
solid surface, it can have   an  atmosphere, basically  hydrogen.  
During  $10$ $Myr$ (an estimated lifetime of  surrounding
stars) the mass $M_\mathrm{atm}$ of the gaseous envelope can  reach
$10 \, M_{\odot}$.  If the density of the atmosphere is $\rho_\mathrm{a}$, and the pressure is $p_\mathrm{a}=2 \, \rho_\mathrm{a} k T/  m_\mathrm{p}$,
where $k$ is the Boltzmann constant, $T$ is the absolute temperature  and $m_\mathrm{p}$ is the proton mass, then 
the height of  the 
homogeneous atmosphere  is
\begin{equation}
h_\mathrm{a}=2 p_\mathrm{a}/ \rho_\mathrm{a} g_\mathrm{m}(R)=  kT/ \
m_\mathrm{p}\ g_\mathrm{m}(R), 
\end{equation}
where  
$g_\mathrm{m}(R)$  is  $g_\mathrm{m}$  at $r=R$.  
  At the temperature
$T=10^{7}\mathrm{K} $  the atmosphere height
$h_\mathrm{a} \sim  10^{8} \mathrm{cm}$.  The atmospheric density
$\rho_\mathrm{a}=M_\mathrm{atm}/4\pi R^{2}h_\mathrm{a}  \sim 10^{3}   \mathrm{g} \ \mathrm{cm}^{-3}$.
Under such a condition,  a  hydrogen combustion 
 must begin already in our time. 

Of course, the accretion rate in the past  could be of
many orders more   if  surrounding stars have been  born in a
molecular cloud which was  located in this region.  (See a discussion  in (\cite{gensel1}; \cite{ghez1}). In this case the combustion could  begin many time    ago and must be more intensive.

A relationship between the temperature and  density can be found from the
thermodynamical equilibrium equations%
\begin{equation}
\frac{1}{4\pi r^{2}\rho_\mathrm{a}}\frac{dL_\mathrm{r}}{dr}=\epsilon_\mathrm{pp}
+\lambda \frac{\dot{M} c^{2}} {M_\mathrm{atm}}-T\frac{dS}{dt},
\end{equation}
where
\begin{equation}
L_\mathrm{r}=4\pi r^{2}\frac{a\, c}{3  \kappa \rho_\mathrm{a}}\frac{d}{dr}T^{4}.
\end{equation}
In these equations
$\dot{M}$ is the rate of  the gas accretion , $\epsilon_\mathrm{pp}$ is the rate of the nuclear energy generation per  unit of mass, $\lambda$ is the portion
of the  energy thermalized at   accretion,
$S$ is entropy, and $a$ is the radiation constant. 
For the proton-proton cycle the value of  $\epsilon_\mathrm{pp}$  can be taken to a first approximation as  
  (\cite{schwar}; \cite{bisnovatyi})
\begin{equation}
\begin{split}
 & \epsilon_\mathrm{pp}=2.5 \, 10^{4} X_\mathrm{H}^{2} \left( \dfrac{T}{10^9}\right) ^{-2/3}  \\   
 &  \times \, \, exp\left[   -3.38 \left( \dfrac{T}{10^9}\right) ^{-1/3} \right] \, \, \, 
  \mathrm{erg}\, \mathrm{g} ^{-1}\, \mathrm{s}^{-1}, 
\end{split}
\end{equation}   
where $X_\mathrm{H}$ is  the hydrogen mass fraction which is assumed to be equal to  $0.7  $, and
 \begin{equation}
\begin{split}
 &    \kappa= 0.2 (1+X_\mathrm{H})  \\ 
 & +    4 \,10^{24}  (1+X_\mathrm{H}) \rho_\mathrm{a}\,  T^{-3.5}  \, \, \,  \mathrm{cm}^{2} \mathrm{g} ^{-1}
\end{split} 
 \end{equation}  
  is  the absorption coefficient caused by the Thomson scattering and  free-free transitions. 
  
In the stationary case for the
homogeneous atmosphere we have%
\begin{equation}
\frac{acT^{4}}{3 \kappa  \rho_\mathrm{a}^{2} h_\mathrm{a}^{2}}=\epsilon_\mathrm{pp}+\lambda\frac{\overset{\cdot}%
{M}c^{2}}{M_\mathrm{atm}}.
\end{equation}

\begin{figure}
\resizebox{\hsize}{!} {\includegraphics[]{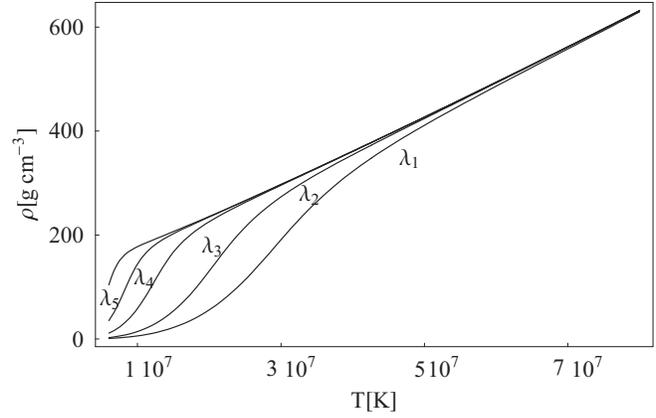}}
 \caption{Relationship between temperature and pressure of the homogeneous
atmosphere for the values of the parameter $\lambda:$ $\lambda_{1}
=1,\lambda_{2}=1/6,\lambda_{3}=10^{-2},\lambda_{4}=10^{-3},\lambda_{5}
=10^{-4}$ at $M=3\,10^6 M_{\odot}$ and $\dot{M}=10^{-7} M_{\odot} \mathrm{yr}^{-1}$}
 \label{fig: RelTP}
\end{figure}

Fig. \ref{fig: RelTP} shows the relationship between $T$ and $\rho_\mathrm{a}$ for
several values  of the parameter $\lambda$.  It follows from the
figure that the hydrogen burning can occur at the density of
$10^{2}\div10^{3}$ $\mathrm{g} \, \mathrm{cm}^ {-3}$.

A  luminosity from the burning layer and  density  distribution  
  can be found by solving
 the differential equations system%
\begin{equation}
\begin{split}
\label{AtmProfile}
& \frac{dM_\mathrm{ra}}{dr} =4\pi r^{2} \rho_\mathrm{a},\ \frac{dp_\mathrm{a}}{dr}=\rho_\mathrm{a} g(R),\\
& \ \frac{dL_\mathrm{r}}{dr}=4\pi
r^{2}\rho_\mathrm{a}\epsilon_\mathrm{pp} , \, 
 \frac{dT}{dr}=-\frac{3\kappa\rho_\mathrm{a}  L_\mathrm{r}}{16\pi\, a\, c\, r^{2}T^{3}},
\end{split}
\end{equation}
where $M_\mathrm{ra}$ is the mass of the spherical layer from the surface to the distance  $r$ from the centre the object. The
boundary conditions  on the surface are of the form: $M_\mathrm{ra}(R)=0,\,  $
$\rho_\mathrm{a} (R)=\rho_{0},$ $T(R)=T_{0},$ $L_\mathrm{r}(R)=0$.  At 
$T_{0}=10^{7}\mathrm{K} $  the luminosity
$L=2.4\, 10^{31}\mathrm{erg}\ \mathrm{s}^{-1}$
at $\rho_{0}=10^{2}\mathrm{g} \ \mathrm{cm}^{-3}$ and $L=5.6 \, 10^{33}\mathrm{erg}\ \mathrm{s}^{-1}$ at $\rho_{0}%
=10^{3}\mathrm{g} \ \mathrm{cm}^{-3}$. 

The temperature at the surface cannot be much more than the above magnitude.  At the  temperature
$T_{0}=3\, 10^{7}\mathrm{K} $ and  $\rho_{0}=10^{3}\, \mathrm{g} \ \mathrm{cm}^{-3}$ the luminosity $L=1.0\, 10^{36}\, \mathrm{erg}\ \mathrm{s}^{-1}$, i.e.  is a  value 
of the order of  the observed bolometric luminosity of Sgr A*.  

\begin{figure}
\resizebox{\hsize}{!} {\includegraphics[]{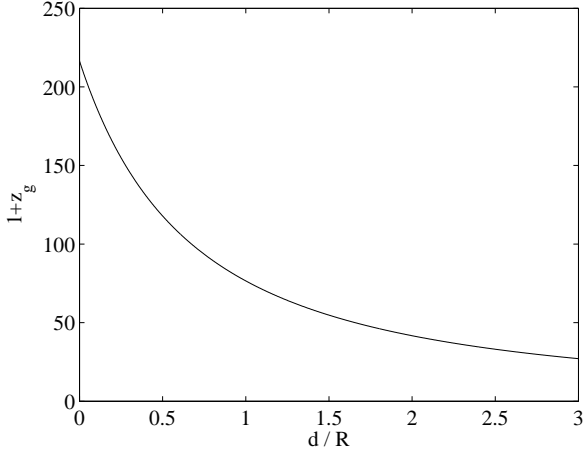}}
 \caption{The gravitational redshift close to the object surface of  the  mass
$M=3 \,10^{6} M_{\odot}$ and  radius $R=0.04 \, r_\mathrm{g}$ as the function of the distance $d$ from the surface ( in  units of $R$). }
 \label{fig: redshift}
\end{figure}

Due to   gravitational redshift a local frequency (as  measured by a local observer)    differs  from the one for a remote observer by the factor  \footnote{
The simpest way to show it is follows. From the viewpoint of a remote observer, 
the energy integral of the motion of a particle in  spherically-symmetric
field is (\cite{verozub91})
\begin{equation}
\dot{r} \frac{\partial \mathcal{L}}{\partial \dot{r}} + \dot{\varphi} \frac{\partial \mathcal{L}}{\partial \dot{\varphi}} - 
\mathcal{L} = E  \nonumber
\label{EnergyIntegral}
\end{equation} 
where $\mathcal{L}$ is the Lagrange function
\begin{equation}
\mathcal{L} = -m c \left(  c^{2} C - A \dot{r}^{2} - B \dot{\varphi}^{2} \right)^{1/2} \nonumber
\end{equation} 
and $E$ is the energy of the particle. In the proper frame of reference of the particle the
above equations take the form:  
$- \mathcal{L} = E$,  and  $\mathcal{L} = -m c^{2} \sqrt{C}$.  
 Photons  can be considered as a particles of the effective proper mass $m_\mathrm{eff} = h \nu / c^{2}$.  It yields  the 
relationship $\nu \sqrt{C} = \nu_{\infty}$, where $\nu_{\infty}$ is the frequency at infinity.}  
\begin{equation}
 (1+z_\mathrm{g})^{-1}=\sqrt{C},
 \label{1+z}
\end{equation} 
where the function  $C=C(r)$   is defined by Eqs. (\ref{ABC}).  Fig. \ref{fig: redshift} shows the dependency
of the redshift factor on the distance $d$   from the surface of the object. 

The difference between  the local and observed frequency can be significance for the  emission emerging from small distances from the surface.  
For example, at  the above luminosity $L=5.6\, 10^{33}\,  \mathrm{erg} \, \mathrm{s}^{-1}$ 
the local frequency of the maximum of the blackbody 
  emission is 
equal to $5.7\, 10^{14} \mathrm{Hz}$. The observed frequency of the maximum  is equal to $2.7 \, 10^{12} \mathrm{Hz}$.   The corresponding specific luminosity  is   
 $ \sim   2\, 10^{21}\, \mathrm{erg} \, \mathrm{s}^{-1} \mathrm{Hz} ^{-1}$.
At  $L=10^{36}\,  \mathrm{erg} \, \mathrm{s}^{-1}$ 
the frequency of the maximum is equal to $2 \,10^{15} \,  \mathrm{Hz} $ ,
the observed frequency is equal to $9.7 \,10^{12} \,  \mathrm{Hz} $, and 
the specific luminosity is  $ \sim   1\, 10^{23}\, \mathrm{erg} \, \mathrm{s}^{-1} \mathrm{Hz} ^{-1}$.
These magnitudes are close to observation data.  ( \cite{melia1};  \cite{falcke2})

\section{ Peculiarity of  accretion  }
The main peculiarity of a  spherical supersonic accretion onto  
 supermassive objects without  event  horizon is the existence of
the second sonic point -- in a vicinity of the object  which is
not connected  with hydrodynamical  effects. The physical reason is
that as the sound velocity $v_\mathrm{s}$ in the incident flow grows
together with the  temperature,  the gas velocity $v$,  like 
the velocity of free falling particles  (\ref{vff}  )    , begin to decrease  close to
$r=r_\mathrm{g}$.  As a result, the equality $v=v_\mathrm{s}$
take place  at some  distance $r_\mathrm{s} > R$,      not far from  the  surface of the object.

According to the used  gravitation equations  (\cite{verozub91}),  the maximal radial velocity of
a  free falling particle does not exceed $0.4\ c$.  Therefore, 
the 
Lorentz-factor is nearly $1$, and to find   $r_\mathrm{s}$     we  can proceed from the simple  hydrodynamics equations:

\begin{equation}
\begin{split}
\label{hydroeqs}
& 4\pi r^{2}v    =\overset{\cdot}{M} ,\ \ \ vv+ {n}^{ \prime   }/n=g\\
& \left( \frac{\varepsilon}{n}\right)^{^{\prime}}-P\frac{n^{\prime}}{n}=0. 
\end{split}
\end{equation}
          
In these equations.  $n$ is the  particles  number density, $\varepsilon$ is the  energy density of the gas,
 $g=F/m$  is the gravitational acceleration. The gas pressure    $P$ is taken as  $2 n k T$,  and the prime denotes
 the derivative with respect to $r$.
  Therefore, as in the case of the Bondi accretion model, we neglect  radiation pressure
,  an equipartition magnetic field
which may exist in the accretion flow, viscosity and    loss of   energy due to   radiation.
However, 
the energy  density of the infalling gas  in  (\ref{hydroeqs})  is calculated more accurate  ( \cite{melia}):
\begin{equation}
\varepsilon=m_\mathrm{p}c^{2}n+\phi nkT,
\end{equation}
where
\begin{equation}
\begin{split}
& \phi=3+x\left(  \frac{3K_{3}(x)+K_{1}(x)}{4K_{2}(x)}-1\right) \\ 
 & +y\left(
\frac{3K_{3}(y)+K_{1}(y)}{4K_{2}(y)}-1\right)  ,
\end{split}
\end{equation}
$x=m_\mathrm{e}c^{2}/kT $, $ y=m_\mathrm{p}c^{2}/kT$ and $K_{j}$ ($j=2,3$) is the $i^{\mathrm{th}}$ order modified Bessel
function, $m_\mathrm{e}$ is the electron mass

 The sound velocity  $v_\mathrm{s}$ as a  function of $r$ can be found    (\cite{service})  as  
\begin{equation}
v_\mathrm{s}=c \left(  \frac{\Gamma P}{\rho\, c^{2}+P}\right) ^{1/2},
\end{equation}
where $\rho=m_\mathrm{p} n$ is the  gas  (fully ionised hydrogen plasma) density.
 The adiabatic  index
\begin{equation}
\Gamma=c_\mathrm{p}/c_{v},
\end{equation} 
where
\begin{equation}
\begin{split}
 & c_\mathrm{p}=5\, \varsigma\,   \Theta^{-1} -(\varsigma^{2}-1) \Theta^{2}), \\   
&  c_{v}= c_\mathrm{p} -1,\\
& \varsigma = K_{3}(\Theta^{-1})/K_{2}( \Theta^{-1}  ),
\end{split}
\end{equation}
and  $\Theta=k T/m_\mathrm{e} c^{2}$.  Then 

\begin{equation}
\frac{P}{\rho c^{2} +P}=\frac{\Theta}{\eta},
 \end{equation}
where   $\eta=K_{3}(\Theta^{-1})/K_{2}(\Theta^{-1} ) $.  Having used
 results by  \cite{Press}  Service     (\cite{service})     have obtained      the following polynomial fitting for
$\Gamma$    and  $P/(\rho c^{2} +P   ) $: 
\begin{equation}
\begin{split}
& P/(\rho c^{2} +P   ) =0.03600 y +0.0584 y^2 - 0.1684 y^3,   \\
& \Gamma   =(5-1.8082 y \\
& +0.8694 y^2 -0.3049 y^3 +0.2437 y^4),  \\
\end{split}
\end{equation} 
where  $y=\Theta/(0.36+\Theta)$. 
We assume  that $\dot{M}=10^{-7} M_{\odot}\ \mathrm{yr}^{-1}$ and that at the distance from the centre
$r_{0}=10^{17}\mathrm{cm}$ the velocity $v(r_{0})=10^{8}\mathrm{cm}\ \mathrm{s}^{-1}$ and the
temperature $T(r_{0})=10^{4}\mathrm{K} $ .  Under  these  conditions $r_\mathrm{s}=8 R$.
At this point the solutions of the equations
have a peculiarity.
   The temperature $T(r_\mathrm{s}) \sim 10^{10} \mathrm{K} $.  The value of  $r_\mathrm{s}$  weakly dependens
    on $\dot{M}$, however, it is more sensitive to    physical conditions
at large $r$. For example, it varies from the above magnitude $8 R$ up to 
   $22 R$  at the temperature  $T(r_{0})=10^{6}\mathrm{K} $ .
 
  The postshock region stretches  from $r_\mathrm{s}$ to  such a depth  where 
    protons stopping occurs due to Coulomb collisions with atmospheric electrons. 
   (We do not take into account  collective effects in  plasma
  ).
 The change in the velocity $v_\mathrm{p}$ of  infalling protons in  plasma is given by the solution of the differential equation
  
  \begin{equation}
  v_\mathrm{p}  \frac{d v_\mathrm{p}}{dr}=g+w,
  \label{dVpdr}
\end{equation} 
where  $w$ is the deceleration of protons due to Coulomb collisions.
This magnitude is  of the form  (\cite{alme};  \cite{Li};  \cite{deufel})
\begin{equation}
w=- f(x_\mathrm{e}) \frac{4 \pi N e^{4}}{m_\mathrm{e} m_\mathrm{p} v_\mathrm{p}} \ln{ \Lambda}.
\label{deceleration}
\end{equation}   
In this equation  $e$ is the electron charge,  $N$ is the particles number  density. At $r<1.3 R  $
the  last magnitude  is taken here as  the quantity 
$n_\mathrm{a}=\rho_\mathrm{a}/m_\mathrm{a}$ resulting from the solution  of Eqs. ( \ref{AtmProfile} ),    and 
as 
  the  particles number density in a free-falling  gas   
 \begin{equation}
n=\dfrac{\dot{M}}{4 \pi r^{2}  v_\mathrm{ff} m_\mathrm{p}},
\end{equation}    
at $r > 1.3 R$,   where  $n > n_\mathrm{a}$.  
The magnitude
\begin{equation}
\ln{ \Lambda} =  \ln{\frac{3}{2 \sqrt{\pi} e^{3}}   \frac{(k T)^{3/2}}{N^{1/2}}  }
\end{equation} 
is the Coulomb logarithm.
 The function $f(x_\mathrm{e})$  is taken here as 
 
  \begin{equation}
f(x_\mathrm{e})=\psi (x_\mathrm{e})-x_\mathrm{e} (1+m_\mathrm{e}/m_\mathrm{p}) \psi' (x_\mathrm{e}) ,
\end{equation} 
where
\begin{equation}
x_\mathrm{e}=\left(  \frac{m_\mathrm{e} v_\mathrm{p}^{2}}{2 k T} \right)^{1/2}, 
\end{equation} 
   $\psi$  is the error function 
  \begin{equation}
\psi(x_\mathrm{e}) = \frac{2}{\sqrt{\pi}} \int_{0}^{x_\mathrm{e}} exp(-z^{2} dz
\end{equation} 
 and $\psi' (x_\mathrm{e})$ is the derivative with respect to $r$.

The  velocity  $v_\mathrm{ff}$  is close to the  greatest possible velocity of the infalling gas.  
 It leads to  the   minimally possible  distance of  a point of the full proton  stopping from the object surface.
 A numerical solution of Eq. ( \ref{dVpdr}  ) shows that this magnitude is equal to
   $( 0.3 \div 0.4)  R$.
  The particles number  density in this region is $\sim 10^{14}\, \mathrm{cm}^{-3}$
  (We neglect 
some difference between the height of protons and electrons stopping (\cite{bildsten})).
  Above this region the most part of the kinetic energy protons is released.

The distribution of the temperature $T(r)$ in this region can be obtained  from an equation of the energy balance.
(Such a question for  nutron stars has been considered  earlier by Zel'dovich \& Shakura (1969) , Alme \& Wilson (1973),
 Turolla et al. (1994), Deufel et al. (2001), and, for 
magnetic white dwarfs, by Woelk \& Beuermann (1992)).   Suppose that all protons stopping power is converted  into radiation. 
  Let $H$ be the radiative flux and $L_\mathrm{r}=4 \pi r^{2} H$  --  the flux trough the sphere of the radius $r$.
Then the energy balance means that 
\begin{equation}
L_\mathrm{r}^{'}=4 \pi r^{2} \epsilon_\mathrm{cul} \, \rho
\label{EquilibrEq0}
\end{equation} 
where $\epsilon_\mathrm{cul}$ is the rate of the energy release due to Coulomb collisions, $\rho= m_\mathrm{p} N$ is the gas 
density and the prime
denotes the derivative with respect to $r$.  The left-hand side    in (\ref{EquilibrEq0})  is
\begin{equation}
L_\mathrm{r}^{'}=4 \pi r^{2} \left( H^{'} +\frac{2}{r} H \right) .
\label{L_Prime}
\end{equation} 
The first-order moment equation of  the standard transfer equation for a spherically- symmetric gray atmosphere
can be written as  ( \cite{schwar}    )
\begin{equation}
H^{'} +\frac{2}{r} H +c \kappa \rho U - j \rho =0,
\label{momentEq  }
\end{equation} 
where $U$ is the density of the radiation energy, $j$ and $\kappa$ are the mean  emission and adsorbtion coefficients,
respectively.

Eqs. (\ref{L_Prime}) and ( \ref{momentEq  }  ) yield
\begin{equation}
L_\mathrm{r}^{'} 4 \pi r^{2} \rho =( j - c \kappa U).
\end{equation}   
Therefore,  the  appropriate for the  purpose of this paper equation of the energy balance is of the form
\begin{equation}
\epsilon_\mathrm{cul} = j -c \kappa U.
\label{EquilibrEq}
\end{equation}
To a first approximation,  $j$  and $\kappa$ can be taken as ( \cite{bisnovatyi})
\begin{equation}
\begin{split}
\label{jrad}
& j=5 \,10^{20}\,  \rho T^{1/2} +6.5\,  U\,  T  \\  
& +2.3\, T B^{2} \, \, \, \mathrm{erg}\,  \mathrm{g} ^{-1} \, \mathrm{s}^{-1}  
\end{split}
\end{equation}
and
\begin{equation}
\begin{split}
\label{kabsorb}
& \kappa=0.4 + 6.4 \,10^{22} \rho\,  T^{-7/2}  \\ 
& +6.5 c^{-1} T_{\gamma} + 2 \,10^{2} B^{2} T^{-3} \,\,\, \mathrm{cm}^{2}\,  \mathrm{g} ^{-1},
\end{split}
\end{equation} 
where $T_{\gamma} =(U/a)^{1/4}$ is the radiation temperature and $B$ is a  possible  magnetic field 
close to the
 object surface  (See Section 5).
Terms in (\ref{jrad}) describe free-free, Compton and synchrotron emission, respectively.  Terms in
( \ref{kabsorb}  ) describe the inverse processes, and  the first term takes into account  the Thomson scattering.

The rate   $\epsilon_\mathrm{cul}$    of the energy release due to  Coulomb collisions  
 per unit of mass is given (\cite{alme}) by
\begin{equation}
\epsilon_\mathrm{cul}=f(x_\mathrm{e}) \frac{4 \pi e^{4}}{m_\mathrm{p} v_\mathrm{p}} \ln{ \Lambda}.
\end{equation} 

If   to  neglect  adsorbtion terms    in  Eq.  ( \ref{EquilibrEq}  )  and  the gravitation influence  
inside   the stopping region   , it is possible to obtain some simple analytic relationships
 between  the 
temperature $T$ and the energy $E_\mathrm{p}$ of a decelerating photon. 
Indeed, the value of $N$ at the distance  $r=6 R$ is of the order of $10^8\, \mathrm{cm}^{-3} $.  At the temperatures $\sim 10^{10} \mathrm{K} $ 
the free-free and Compton radiations are predominated.
Taking into account  that
  $x_\mathrm{e} \ll 1$, we obtain
\begin{equation}
f(x_\mathrm{e}) \approx \frac{4}{3 \sqrt{\pi   }} \left( \frac{m_\mathrm{e}}{m_\mathrm{p}}\right) ^{3/2}  \left( \frac{E_\mathrm{p}}{k T}\right) ^{3/2},
\end{equation}
which yields the  rate  of the energy release per unit of mass: 
\begin{equation}
\epsilon_\mathrm{cul}=\xi \frac{N E_\mathrm{p}}{T^{3/2}},
\end{equation} 
where
\begin{equation}
\xi=\frac{16 \sqrt{\pi m_\mathrm{e}} e^{4} }{3 \sqrt{ 2 } m_\mathrm{p}^{2} k^{3/2}}.
\end{equation} 
Then, if the
 largest contribution provides  free-free radiation  ( the field $B$ is  weak )  ,  it follows from the equation 
of the energy balance 
that 
\begin{equation}
T=   \left( \delta_{1} E_\mathrm{p}\right) ^{1/2}
\end{equation} 
where $\delta_{1}=2.36 \,10^{24}\, \mathrm{K} ^{2} \, \mathrm{erg}^{-1}   $. 
For example, at $r=6 R$ the proton energy  $E_\mathrm{p}=4.8 \,10^{-6} \, \mathrm{erg}$ which
yields the tempereture $T=3.3 \,10^{9}\mathrm{K}  $.  Since  $E_\mathrm{p}$  is a fast decreasing function 
close to  
$r= 1.4 R$, a sharp drop in the temperature takes  place at the  distance   $\sim 0.4 R$ from the centre.

If the magnetic field is sufficiently strong,  the main contribution in the energy balance  provides  synchrotron radiation.
In this case the relationship between
the temperature    $T$ and the energy  $E_\mathrm{p}$   of the proton   takes the form
\begin{equation}
T=\left(  \delta_{2} \frac{N E_\mathrm{p}}{B^{2}} \right)^{2/5} 
\end{equation} 
where $\delta_{2}=10^{24} \mathrm{K} ^{5/2} \,  \mathrm{erg}^{-1} \mathrm{cm}^{3}$
and $B\neq 0$.

In general case, assuming that at the distance $r_{0}=6 R$  the luminosity $L(r_{0})=10^{36} \mathrm{erg} \, \mathrm{s}^{-1}$ , the energy flux
$U(r_{0})=L(r_{0}) /4 \pi r_{0}^{2} c$ and the protons velocity
$v_\mathrm{p}(r_{0})=v_\mathrm{ff}(r_{0})$, we find from Eq. (\ref{EquilibrEq}) that 
the temperature $T(r_{0})=5.4 \,10^{9} \mathrm{K} $ . 

On the other hand, if the absorption is not ignored,  Eq. (\ref{EquilibrEq}) yields  the differential equation
\begin{equation}
T^{\prime}=-\frac{\partial  \Phi(r,T) / \partial r}{\partial \Phi(r,T) / \partial T} ,
\label{Tprime}
\end{equation}  
where  $\Phi(r,T) =\epsilon_\mathrm{cul} - j + c \kappa U$. 
A simultaneously solving of this differential equation together with the equations 
\begin{equation}
\frac{1}{3} U' = \kappa  \frac{L_\mathrm{r}}{4 \pi r^{2} c} \\ \nonumber
\label{Uprime}
\end{equation} 
and  (\ref{EquilibrEq0}) 
 allows to obtain the distribution $T(r)$ at   the  distances $r$  where 
 $\epsilon_\mathrm{cul} \neq  0$.

\begin{figure}
\resizebox{\hsize}{!} {\includegraphics[]{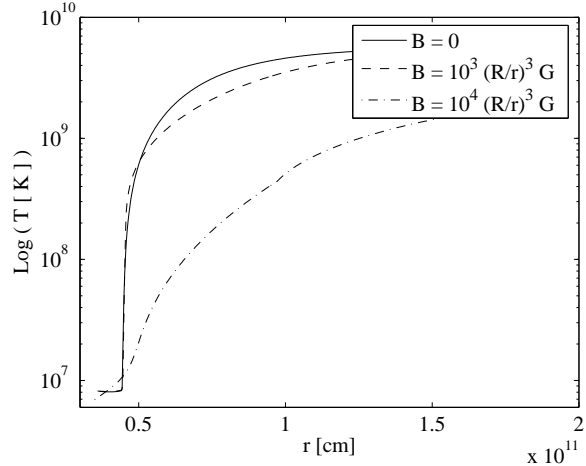}}
 \caption{The temperature $T$ as the function of   $r$  inside the region  of protons stopping for the object of the radius
$R=0.04 r_{g}$.
 The boundary conditions are:  at the distance from the surface $5 R$ the luminosity $L_{0}=10^{36}\mathrm{erg}\ \mathrm{s}^{-1}$, the temperature
  $T_{0}=5.4 \,10^{9}\mathrm{K} $,  the velocity of protons  $v_{p0}=2.32 \,10^9 \mathrm{cm} \, \mathrm{s}^{-1}$.
  The plots for three value of the magnetic field $B$ are shown.
 } 
 \label{fig: JumpT}
\end{figure}

Fig. \ref{fig: JumpT} shows a typical temperature profile in the region of  protons stopping.

It is interesting to estimate the temperature at the surface proceeding from the simplifying assumption that
all  luminosity of Sgr A* ($L \sim 10^{36} \mathrm{erg}\, \mathrm{s}^{-1}$) is caused by nuclear burning at the surface and stopping of incident protons at some distance  $d\leq R$ from the surface. Under such a condition, 
     an effective temperature $T_\mathrm{0}$  on the top of the dense atmosphere ( $d \sim 0.4 R$ )     is about
 $10^{4} K  $.
The radial distribution of the temperature in the diffusion approximation can be  determined by the differential 
equation  \begin{equation}
 (T^{4})^{'} = \frac{3 \chi}{4 b_\mathrm{s}} H, 
\end{equation} 
where $\chi=\kappa \rho$,  $b_\mathrm{s}= 5.75 \, 10^{-5} \mathrm{erg}\,  \mathrm{cm}^{-2} \mathrm{K}^{-4} $ is the Stephan-Boltzmann constant,
 and
$H$ is the radiative flux which is supposed to  constant.
It gives approximately
\begin{equation}
T^{4} - T_{0}^{4} = \frac{\chi}{b} H \Delta r,
\end{equation} 
where  $\Delta r= 0.4 R$ . Setting $\chi=1$ we find that at the surface
the temperature is about  $ 10^{7}\mathrm{K} $ which  coincides  with the magnitude used in  Section 2 . 

To find  the temperature distribution  from the surface to distances  $d \sim R$ more correctly,  equations 
of the structure of  lower atmosphere,  accretion and  energy balance 
 should be solved  simultaneously. It can affect the value of the atmospheric density at the surface. However, as it was noted in Section 2, our knowledge of this magnitude initially is rather uncertain. The density of $\rho \sim 10^{3} \mathrm{g} \mathrm{cm^{-3}}$ (which corresponds to the temperature $T \sim 10^{7} \mathrm{K}$) is the most probable since an essentially greater value of the density or temperature of the bottom atmosphere leads to the luminosity which contradicts  the observable bolometric luminosity of  Sgr A*. At the same time the existence of a high temperature at distances where protons stopping occurs  hardly is a subject to doubt. Therefore,  the  jump of the temperature of the order of $(10^2 \div 10^{3} ) \mathrm{K}$ is an inevitable consequence of the Sgr A* model under consideration. 

\section{ Comptonization  } An emergent intensity $I$ of the
low-atmospheric emission after passage through a hot spherical
homogeneous layer of gas is a convolution   of the Plankian
intensity $I_{0}$  of  the emission of a  burning layer, 
 and the frequency redistribution
function $\Phi (s)$ of $s=\lg(\nu/\nu_{0})$,  where $\nu_{0}$ and
$\nu$ are the frequencies of the
incoming and emerging radiation, respectively:%

\begin{equation}
I(x)=\int_{-\infty}^{\infty}\Phi(s)I_{0}(s)ds.
\end{equation}
 At the temperature of
the order of $10^{10}\mathrm{K} $ the dimensionless parameter
$\Theta=kT/m_\mathrm{e}c^{2}\sim1$. Under the
condition the function $\Phi(s)$ can be calculated
(\cite{brinkshaw}) as follows:

\begin{equation}
\Phi(s)=\sum_{k=0}^{m}\frac{e^{-\tau}\tau^{k}}{k!}P_{k}(s),
\end{equation}
where
\begin{equation}
P_{1}(s)=\int_{(\beta_\mathrm{e})_{\min}}^{1}
\varphi(\beta_\mathrm{e}) P(s,\beta_\mathrm{e})d\beta_\mathrm{e}.
\end{equation}
In the last equations  $\beta_\mathrm{e}$ is the electron-light velocity speed ratio,
\begin{equation}
\varphi(\beta_\mathrm{e})=\frac{\gamma^{5} \beta^{2} exp( -\gamma/\Theta )}{\Theta K_{\mathrm{2}}( \Theta^{-1} )},
\end{equation}  
$\gamma=(1-\beta_\mathrm{e}^{2})^{-1/2}.$, 
\begin{equation}
(\beta_\mathrm{e})_{\min}=\frac{e^{\shortmid s\shortmid}-1}{e^{\shortmid s\shortmid}+1}%
\end{equation}

and
\begin{equation}
\begin{split}
& P(s,\beta_\mathrm{e})=\frac{3}{16\gamma^{4}\beta_\mathrm{e}}\int_{\mu_{1}}^{\mu_{2}}(1-\beta_\mathrm{e}\mu)^{-3} \\  
& \times  \left(
1+\beta_\mathrm{e}\mu^{\prime})(1+\mu^{2}\mu^{\prime2}
+\frac{1}{2}(1-\mu^{2})(1-\mu^{\prime2})\right)
 d\mu, \nonumber
\end{split}
\end{equation}%

\begin{equation}
\mu^{\prime}=\frac{e^{s}(1-\beta_\mathrm{e}\mu)-1}{\beta_\mathrm{e}},
\end{equation}%

\begin{equation}
\mu_{1}=\left\{
\begin{array}
[c]{c}%
-1\hspace{0.5cm}\hspace{0.5cm}\hspace{0.2cm}\hspace{0.2cm}s\leq0\\
\frac{1-e^{-s}(1+\beta_\mathrm{e})}{\beta_\mathrm{e}}\hspace{0.5cm}s\geq0
\end{array}
\right.
\end{equation}%

\begin{equation}
\mu_{2}=\left\{
\begin{array}
[c]{c}%
\frac{1-e^{-s}(1-\beta_\mathrm{e})}{\beta_\mathrm{e}}\hspace{0.5cm}\hspace{0.5cm}\hspace
{0.2cm}\hspace{0.2cm}s\leq0\\
1\hspace{3 cm}s\geq0.
\end{array}
\right.
\end{equation}

\begin{figure}[h]
\resizebox{\hsize}{!} {\includegraphics[]{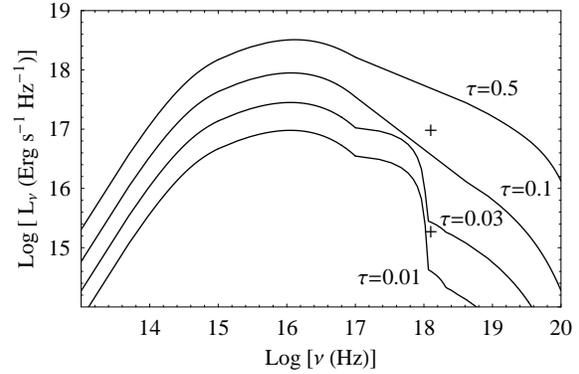}}
 \caption{ Emergent spectra of    nuclear burning after comptonization by a  homogeneous     hot layer for
 the optical thickness $\tau=0.01,\, 0.03\, ,0.1,\, 0,5$ . The crosses denote the observation data  (\cite{baganoff}; \cite{ porquet}) (The upper and lower limits are specified).}
  \label{fig: ComptonSpectrum}
\end{figure}

Taking into account scattering
 up to 3rd order ( $m=3)$   an approximate emergent spectrum
of the nuclear radiation   after
comptonization  has been obtained for the case when
  the  temperature at the surface is $10^7 \mathrm{K}$ and $\rho=3\, 10^{3} \mathrm{g} \mathrm{cm}^{-3}$.
   The results are plotted in Fig. \ref{fig: ComptonSpectrum} for several values of the optical thickness $\tau$ .

It follows from this  figure that   the
comptonization  may  give a contribution to the observed Sgr A* radiation
(\cite{baganoff}; \cite{ porquet})  in the range $10^{15} \div 10^{20}\, Hz$ .

It is necessary to note that the
 timescale $\Delta t$ of a variable process is connected with the size $\Delta $
 of its region by the relationship $ \Delta \leq c\ \Delta t/(1+z_\mathrm{g})$.
 For example,  the variations
 in the radiation intensity of $\sim 600\ s$ on  clock of  a remote observer occur    at the
distance $d \leqslant 1.7 R$ from the surface. For this reason high-energy flashes 
   can be
interpreted as processes near the surface of the central objects.

It is possible that the diffuse, hard X-ray emission from the
giant molecular clouds of  central $100 pc$ of our Galaxy  (\cite{park}; \cite{ sunyaev})    is a
consequence of the  more intensive X-ray emission of Sgr A* due to nuclear bursts  in the past .

\section{Transfer equation and synchrotron radiation of Sgr A* }

Some authors (\cite{robertson})  find in spectra of
candidates into  Black Holes some evidences for the existence of  a  proper magnetic
field of these objects that is incompatible with the existence of an 
event horizon. In order to show how  the 
magnetic field near the surface of the object in question can manifests  itself,  we
calculate the contribution  of a magnetic field to the spectrum of  the synchrotron
radiation.  We  assume  that the magnetic field   is of the form 
\begin{equation}
B_\mathrm{int}=B_{0}  \left( \frac{R}{r}\right)^{3},
\label{Bfield}
\end{equation}
where $B_{0}=1.5\,10^{4}\  \mathrm{G}$.
 
 To find the emission spectrum from the  surface vicinity,
 the influence of
gravitation on the frequency of  photons and their motion  must be
taken into account.  An appropriate transfer equation is the Boltzmann equation for photons which uses 
 equations of their  motion in strong gravitational fields
resulting from  the used gravitation equations. (Such an equation in General Relativity has been  considered  
 by Lindquist  (1966), Schmidt-Burgk (1978), Zane et al. (1996) and Papathanassiou \& Psaltis (2000) )
 The Boltzmann equation is of the form
\begin{equation}
\frac{d\mathcal{F}}{dt}=St(\mathcal{F}, t, \mathbf{x} ), 
\label{Bolz1}%
\end{equation}
where $\mathcal{F}= \mathcal{F}(t, \mathbf{x}, \mathbf{p})$ is the phase-space distribution function, 
$\mathbf{ x }=(x^{1}, x^{2}, x^{3})$ and 
 $\mathbf{ p}=(p^{1}, p^{2}, p^{3})$ are the photon coordinates and momentums, $d/dt$ is the 
 derivative along the photon path,  $ St(\mathcal{F}, t, \mathbf{x} )$  is a collisions integral. 
 The solution of this equation  is related to the  intensity $I$ as 
 $\mathcal{F}=\beta^{-1} I $ 
 where $\beta = 2 h^{4} \nu^{3} c^{-2}$ and $\nu$ is the local frequency. 
 
 In the  stationary spherically-symmetric field the distribution function $\mathcal{F} = \mathcal{F}(r, \mu, \nu)$ where
 $\mu=\cos{\theta}$ is the cosine between the direction of the photon and the surface normal. 
 The path derivative takes the form
 \begin{equation}
\frac{d \mathcal{F}}{dt} = v_\mathrm{ph}  \left(  \frac{\partial \mathcal{F}} {\partial r}  +
 \frac{\partial \mathcal{F}}{\partial \nu}  \frac{\partial \nu}{\partial{r} }  +
  \frac{\partial \mathcal{F}}{\partial \mu}  \frac{\partial \mu}{\partial{r} }      \right) ,
\end{equation} 
where the radial component of the photon velocity $dr/dt=v_\mathrm{ph}$ is given by  (\cite{verozub95}   )

\begin{equation}
v_\mathrm{ph} = c \left[ \frac{C}{A}  \left( 1-\frac{C b^{2}}{f^2}  \right)       \right] ,
\label{photonvelocity}
\end{equation}  
and  $b$ is an impact parameter. 
The frequency as a function of $r$ is given  by the equation $\nu = \sqrt{C}\,  \nu_{\infty} $ .
  The function $\mu =\mu(r)$  satisfy  
 the   differential equation
\begin{equation}
\frac{d\mu}{dr} = (1- \mu^{2})^{2} \frac{d\theta}{dr},
\label{mu}
\end{equation} 
where  (\cite{verozub95})
\begin{equation}
\frac{d\theta}{dr} = \frac{c C b}{f^{2}}.
\end {equation} 
In the above equations  the geometrical relationship  $b^{2} = r^{2} (1-\mu^{2}) $ must be taken into account.
 
With regards to  all stated above, the Boltzmann equation along the photon paths can be taken in the form
\begin{equation}
v_\mathrm{ph} \frac{d\mathcal{F}}{dr} =  \frac{1}{\beta} (\eta+\sigma J)- (\chi +\sigma) \mathcal{F}.
\label{MainTransferEq}
\end{equation} 
The right-hand side of this equation is the ordinary simplified collisions integral   for an isotropic scattering
( \cite{mihalas}      ) in terms of the function $\mathcal{F}$.
In this equation 
 \begin{equation}
J=(1/2)    \int_{-1}^{1} \beta \mathcal{F} d\mu , 
\end{equation} 
is the mean intensity,
  $\eta$ is the synchrotron emissivity, $\chi$ is the true adsorbtion
opacity ,  $\sigma$ is the Thomson scattering opacity. 
 
 If $\mathcal{F}_{\nu}(r, \mu)$ is a solution of  the integro-differential equation ( \ref{MainTransferEq}  ), the 
specific radiative  flux is 
\begin{equation}
H_{\nu} = (1/2) \int_{-1}^{1}\beta \mathcal{F}_{\nu}(r, \mu ) \mu d\mu
\end{equation} 
Therefore, by solving  Eq. ( \ref{MainTransferEq}  ) for all parameters
$0 < b < \infty$ an emergent spectrum of  Sgr A* can be obtain. 
For a numerical solving  an iterative method can be used  with no 
scattering as a starting point.

We assume that  at the surface  of the object  (due to nonstationary MHD-processes)    and  also higher, in 
 the shock  region,  a small  fractions of the electrons  may exist  in a non-thermal distribution.   Therefore, following 
 Yuan, Quatert \& Narayan (2004),  we setting $\eta=\eta_{\mathrm{th}}+\eta_{\mathrm{nth}}$ and $\chi=\chi_{\mathrm{th}}+\chi_{\mathrm{nth}}$, where indexes $\mathrm{th}$ and $\mathrm{nth}$ refer to themal 
and nonthemal population,  respectively. 
 
 There is  a  fitting function for the synchrotron emissivity of the thermal distribution of electrons 
 which is valid from $T\sim 10^{8} \mathrm{K} $  to  relativistic 
 regimes  ( \cite{mni96}  ):
 \begin{equation}
\eta_{\mathrm{th}} = \frac{N e^{2} \nu}{\sqrt{3} c K_{2}(1/\Theta)} \mathcal{M}(\zeta),
\end{equation} 
where,  $K_{2}(1/\Theta)$ is the modified Bessel  function of the second order,
\begin{equation}
\mathcal{M}(\zeta) = \frac{4 q_{1}}{\zeta^{1/6}} \left( 1+\frac{0.4 q_{2}}{\zeta^{1/4}} +\frac{0.53 q_{3}}{\zeta^{1/2}}\right)  e^{-1.8896 \zeta^{1/3}} ,
\end{equation} 
$\zeta=2 \nu /3 \nu_{b} \Theta^{2}$ and $\nu_{b}=e B/2 \pi m_\mathrm{e} c$ is the nonrelativistic cyclotron frequency.
The parameters $q_{1}$, $q_{2}$ and $q_{3}$ are   functions of the temperature $T$. They  have been obtained from
 Table 1 of  the above paper.
 The true adsorbtion opacity $\chi_{\mathrm{th}}$  is supposed to be related to  $\eta_{\mathrm{th}}$ via Kirchoff law as $\chi_{\mathrm{th}} =\eta_{\mathrm{th}}/B_{\nu}$, where $B_{\nu}$ is the Planck   function.
 
 For the  population of non-themal electrons we  use a power-law distribution ( \cite{ozel} ) which is proportional to 
 $ \gamma^{-\alpha}$  where $\gamma$ is the  Lorentz factor and the parameter $\alpha$ is supposed to be equal to 2.5.
 For such a distribution the emissivity of the non-thermal electrons is approximately given by 
 \begin{equation}
\eta_{\mathrm{nth}}=0.53\,  \delta\,   \frac{e^{2} N}{c} \varphi_{1} (\Theta) \nu_{b} \left( \frac{\nu}{\nu_{b}}\right) ^{ -0.25  },
\end{equation} 
 where
 \begin{equation}
\varphi_{1} (\Theta)= \Theta \frac{6+15 \Theta}{4+5 \Theta}
\end{equation} 
 and  the fraction $ \delta $ of the electrons energy   in the thermal distribution is supposed 
 to be the constant in radius and equal to 0.05.
 
 The adsorbtion coefficient for non-thermal electrons is taken as
   \begin{equation}
\chi_{\mathrm{nth}}=1.37\,10^{27} \delta  \frac{e^{2} N}{c\,  \nu} \varphi (\theta) \nu_{b} \left( \frac{\nu}{\nu_{b}}\right) ^{ 2.75 }.
\end{equation} 

\begin{figure}
\resizebox{\hsize}{!} {\includegraphics[]{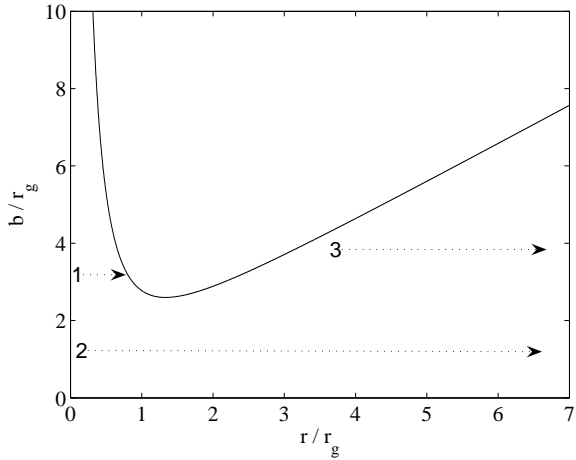}}
 \caption{The function $b(r)$.  There are 3 types of photons trajectories:  1)  The impact parameter $b>b_\mathrm{cr}$, the trajectory begins  at  $r < r_\mathrm{cr}$, 2) $b<b_\mathrm{cr}$ , 3) $b>b_\mathrm{cr}$,  the trajectory begins  at $r>$ $r_\mathrm{cr}$.}
 \label{fig: bPlot}
\end{figure}

For a  correct solution of Eq. (\ref{MainTransferEq}) is
necessary to take into account  that from the point of view of the used
gravitational equations there are three types of photons trajectories in the
spherically-symmetric gravitation field. It can be seen in Fig.
\ref{fig: bPlot}.  The figure  shows the  locus  in which radial  photon
velocities are equal to zero. It is  given by the equation
\begin{equation}
b=\frac{f}{\sqrt{C}}. \label{FotonLocus}%
\end{equation}

The minimal value    of $b$ is   $b_\mathrm{cr}=3\sqrt{3}r_\mathrm{g}/2$. The
correspondsing  distance from the centre is
$r_\mathrm{cr}=\sqrt[3]{19}r_\mathrm{g}/2.$ 
The photons with an impact parameter $b>b_\mathrm{cr}$ 
, the trajectories of which start  at  $r < r_\mathrm{cr}$     ( type 1 in Fig. \ref{fig: bPlot}),
  cannot  move  to infinity. It can do   photons  with
 $ b > b_\mathrm{cr}$,  if their trajectories start at   $r> r_\mathrm{cr}$ ( type 3)    , and all  photons with   $b<b_\mathrm{cr}$  ( type 2).

In order to find the solutions  $\mathcal{F}_{+}$    of equation (\ref{MainTransferEq}),  describing the outgoing emission 
provided by  photons of type 2, 
  the boundary condition $\mathcal{F}(r=R, \nu, \mu) = 0$ can be used. The boundary condition $\mathcal{F}(r \rightarrow  \infty, \nu, \mu) = 0$  allows to  find the solutions   $\mathcal{F}_{-}$    ,  describing the emission ingoing from infinity. The values  of this solution   can be used as a boundary conditions to find  the outgoing emission 
 provided by the photons of type 3 (Zane et al. 1996). 
 Of course, the value of the function $\mathcal{F} = \beta^{-1} B_\mathrm{\nu} $, where  $B_\mathrm{\nu}$ is the Planck  function, can be used as
 a boundary condition to find emission from large optical depths.
 
\begin{figure}[h]
\resizebox{\hsize}{!} {\includegraphics[]{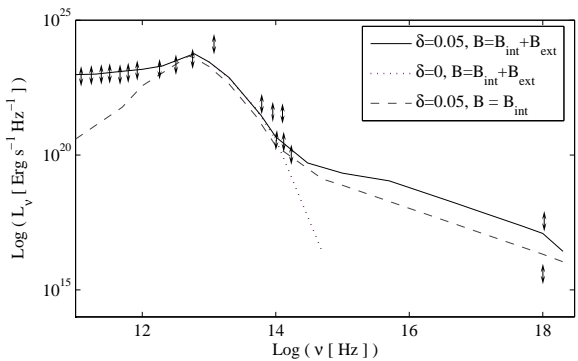}}
\caption{The spectrum of the synchrotron radiation. 
  The dashed line shows the luminosity $L_{\nu}$ for a remote observer for the
  possible proper magnetic field
  $B_\mathrm{int} $.  The parameter $\delta$ of the  hybrid distribution of electrons is equal to  $0.05$.
 The dotted line shows the  luminosity  without non-themal electrons. 
  The solid line is the luminosity at the presence  both  $B_\mathrm{int} $ and 
  the  external  equipartition magnetic field $B_\mathrm{ext}$.
 The short lines with the double arrows show observation data according to  \"Ozel et al (2000)  }
 \label{fig: spectrumS}
\end{figure}

Fig. \ref{fig: spectrumS} shows the spectrum of the synchrotron
radiation in the band of $10^{11}\div2\, 10^{18}\ \mathrm{Hz} $ for three
cases:\\
-- for the   magnetic field   $B_\mathrm{int} $        ( \ref{Bfield})   of
the object at   the  gibrid distribution of electrons with the parameter  $\delta=0.05$ 
(the dashed line),\\
-- for   the sum of      $B_\mathrm{int} $   and  an external equipartition magnetic field       $B_\mathrm{ext}=(\dot{M} v_\mathrm{ff}  r^{-2})^{1/2}$           
  which may exist in the
accretion flow (\cite{melia}) 
 ( the solid line),\\
--  for the  magnetic field  $B_\mathrm{int} $ without non-themal electrons  (the dotted line which at the frequencies          $ < 10^{14}\, \mathrm{Hz} $               coincides with the dash line).
 in the presence of  the proper magnetic field 

It follows from the figure  that the proper magnetic field can manifest itself for the remote observer at the frequencies 
larger that $(2\div4) 10^{12}\,  \mathrm{Hz} $  -- in IR and X-ray radiation. 

It follows from Figs. 5 and 7 that  the perculiarity of   Sgr A* spectrum  at  $\nu  >  10^{14}\, Hz$
can be explained by  both   physical processes at
the object surface  and by the existence of a non-thermal fraction of  electrons in the gas environment.
 We do not know at the moment  which of these sources is predominant.  However, the opportunity of the explanation of  flare activity of Sgr A* by  physical processes at the surface  is rather attractive.

The measurements of the polarisation of Sgr A* emission offer new possibilities for  testing of  models of this object. According to  observations by Bower et al. (2003) at the frequency  $2.3\, 10^{11}\, Hz$,  the Faraday rotation measure (RM) is $\leq 2\, 10^{6}\, rad\, m^{-2}$.  In our case at  this frequency   the optical depth $\tau$ of the atmosphere   become equal to  $1$ at a distance of 
$r_\mathrm{min} \approx 7\, r_\mathrm{g}$ from the centre.  The influence of the internal magnetic field is insignificant at here.  
The value of the RM is $\sim 10^7\, \mathrm{rad}\, \mathrm{m}^{-2}$. 
This magnitude is not differed from that obtained in the model  by Yan, Eliot \& Narayan (2004)   based on the SMBH - theory. Taking into account that we know the distribution of the density and magnetic field very approximately, we cannot say that the obtained  value of the RM  
contradict observation data. 

At increasing of the frequency the value of $r_\mathrm{min}$  decreases, and  influence of the proper magnetic field must increase because the radiation come from more depth of the atmoshere.  At frequencies more than  $2\, 10^{12}\,  Hz$ the atmosphere become the  transparent completely to the synchrotron radiation. Therefore, the full  depolalization of the radiation   may give some evidence for the existence of a magnetic field close the surface.

\section{Conclusion}
The clarification  of  nature of compact objects in the galactic
centres is one of actual problems of fundamental physics and
astrophysics. The results obtained above, of course, yet do not
allow to draw the well-defined conclusions. However they show that
the possibility, investigated here, at last,does not contradict
the existing observations, and require  further study.

\end{document}